# From Complexity Measurement to Holistic Quality Evaluation for Automotive Software Development

Position Paper

Jens Heidrich, Michael Kläs, Andreas Morgenstern, Pablo Oliveira Antonino, Adam Trendowicz
Fraunhofer IESE, Fraunhofer-Platz 1, 67663 Kaiserslautern
*{Jens.Heidrich|Michael.Klaes|Andreas.Morgenstern|*
*Pablo.Antonino|Adam.Trendowicz}@iese.fraunhofer.de*

Jochen Quante, Thomas Grundler
Robert Bosch GmbH, Robert-Bosch-Platz 1, 70839 Gerlingen-Schillerhöhe
*{Jochen.Quante|Thomas.Grundler}@de.bosch.com*

## Management Summary

In recent years, the role and importance of software in the automotive domain has changed dramatically. More functionalities are implemented in software and are enabled by software (such as automated driving functions). Furthermore, vehicles get connected with other vehicles and digital services operated by digital platforms. In order to cope with the increased size and complexity of software, practices from the classical field of software engineering have been increasingly adopted and service-oriented and standardized architectures, like AUTOSAR, have been introduced. One implication of architecture-centric practices is also that more and more code is generated from models, which changes the way software is produced, maintained, and evaluated regarding its quality. Additionally, software development processes have become more and more agile and continuous in order to facilitate faster and incremental implementation and deployment of new functions. BizDevOps reduces the distance between business, development, and operations, enabling practices like over-the-air delivery of new functionalities, improvements, and bug fixes.

Because of these trends, being able to systematically evaluate and manage software quality becomes even more crucial. Measurement-based approaches are the common best practice for doing this. In research, several new approaches have emerged in recent years to address these software trends. In practice, however, we still find a largely static approach for measuring software quality in general and its complexity in particular. For instance, quite commonly, manufacturers require their suppliers to deliver a predefined list of complexity measures with static thresholds to fulfill. This approach neither takes into account that software quality is dependent on the specific context of a function, nor does it address the trends sketched above. We therefore propose evolving the paradigm of measurement-based software quality evaluation and management accordingly. Our proposed approach is holistic and includes consolidated conclusions from recent research as well as experiences from other industrial application fields.

Our proposed framework includes four areas. They define the term software quality in a goal-oriented way for a specific context and describe how to measure and evaluate relevant qualities.

1. Understand the context and the stakeholder goals/needs
2. Derive technical quality factors for relevant development artifacts related to stakeholder goals
3. Determine quality measures and evaluation rules for relevant quality factors
4. Define a process for managing quality during the entire lifecycle

We propose that all parties should agree on goals, factors, measures, evaluation rules, and quality management processes at the beginning of a software development project (this could, e.g., be documented in the specification) and should revise this agreement in predefined cycles based on the course of the project. This framework provides the following advantages:

- Software quality has a clear link to the value it provides to stakeholders in specific contexts
- Only relevant quality factors and measures are included, which reduces overhead
- Application- and goal-oriented evaluation rules avoid comparing apples and oranges

- Focus on all relevant development artifacts (such as architecture and models) and not only on code
- Provides a real value for collecting measurement data during the product lifecycle
- Keeping up with increasing development speed and flexible feature development

# 1 Introduction

In recent years, the role and importance of software in the automotive domain has changed dramatically. Software and its quality have become a major driver of innovation and a differentiator from competitors. This change has been mainly driven by implementing existing functionalities in software, which used to rely on electromechanical components in the past, and also by adding new software-based functionalities such as automated driving functions and integration of the vehicle with digital platforms and ecosystems in which the vehicle becomes part of a digital service.

In order to cope with the increased size and complexity of software-based functions in a vehicle, practices from the classical field of software engineering have been increasingly adopted and more powerful controllers for operating the software have been introduced. This goes hand-in-hand with the paradigm shift towards service orientation and standardized architectures like AUTOSAR. One of the consequences of the wide adoption of architecture-centric practices is that more and more code is generated from models, which changes the way software is produced, maintained, and evaluated regarding its quality.

Additionally, software development processes have also changed radically in the last few years. Processes have become more and more agile and continuous in order to facilitate faster and incremental implementation and deployment of new functions. This paradigm shift is supported by sophisticated Continuous Development and Integration tooling and pipelines. Along these lines, trends towards BizDevOps have been considered as a means to reduce the distance between business, development, and operations, enabling practices like over-the-air delivery of new functionalities, improvements, and bug fixes.

Because of the increased importance of software, being able to systematically evaluate and manage its quality is a crucial success factor. ISO/IEC 25000 [1] describes recommendations for Software Product Quality Requirements and Evaluation (SQuaRE) and defines key software qualities (such as functional suitability, reliability, or maintainability). There are many different methods, techniques, and tools available for software quality assurance. However, for systematically determining, evaluating, and managing software quality in general and its complexity in particular, measurement-based approaches are the common best practice. Measurement is the process by which numbers or symbols are mapped to attributes of entities in the real world in such a way as to describe them according to clearly defined rules [2]. In every engineering discipline, measurement is a core activity and a symbol for mature development processes. Consequently, common maturity and capability models such as SPICE [3–6] define process areas and practices related to software measurement.

In automotive software development practice, we still find a largely static approach for measuring software quality. For instance, quite commonly, manufacturers demand that their suppliers deliver a list of complexity measures [7] with static thresholds to fulfill for the delivered source code. Of course, complexity has been acknowledged to be one of the main factors impacting software quality (e.g., [8]). However, it is only one aspect of many impacting software quality. Furthermore, focusing on source code only in the context of measurement-based quality evaluation is too narrow considering the trends sketched about, e.g., towards more complex software architectures and model-based software development. Furthermore, static approaches also assume that there is a universal generic understanding of which metrics to collect for all kinds of systems and that universal thresholds exist that define what good-quality software systems must look like. This is unfortunately not true, as software quality may be interpreted in different ways for different kinds of systems and stakeholders.

Research in measurement-based software quality evaluation and management has developed many approaches over the years to cope with these deficiencies. Comprehensive solutions for software quality modeling (such as [9]) provide methods, techniques, and tools for systematically specifying how to measure and evaluate different aspects of software quality. We also find more recent approaches, focusing, e.g., on measuring related concepts such as technical debt [10], on dedicated development artifacts such as the software architecture [11], or on dedicated types of systems such as the quality of AI systems [12].

© Fraunhofer IESE and Robert Bosch GmbH 2021    2/21    27.10.2021

We therefore propose changing the paradigm of measurement-based software quality evaluation and management in practice to a more modern and holistic approach, including conclusions from recent research and addressing novel technologies and development processes.

This paper is organized as follows: In Section 2, we take a deeper look at relevant trends in automotive software development that have an impact on software quality and how to evaluate and measure it. In Section 3, we will discuss the current state of the practice of measurement-based complexity evaluation in automotive software development. Section 4 will highlight some relevant approaches from research in software complexity and quality evaluation. Based on that, in Section 5 we will summarize the main needs for action for measurement-based software quality evaluation and management from our point of view. In Section 6, we will propose a quality evaluation framework for automotive software development addressing the trends and progression in research. Section 7 summarizes the advantages of the proposed framework and illustrates next steps towards holistic quality evaluation for automotive software development.

## 2 Relevant Trends in Automotive Software Development

Several trends are changing automotive software development. In this section, we will summarize the trends that are most important for software quality and complexity evaluation. We will shed light on the change of paradigms, like the shift from hand-written code to generated source code, the change of processes from a classical V-model to more agile processes like DevOps or Continuous Deployment, but also on the introduction of disruptive technologies, especially in the area of Artificial Intelligence, which are relevant in the context of, e.g., environmental perception for highly automated and autonomous cars.

The introduction of highly automated driving assistance functionalities like automatic cruise control systems in modern cars led to a massive increase in the complexity of the control algorithms for the physical parts of a vehicle. This fostered a paradigm change from manual coding to model-based design with tools like Matlab/Simulink, where algorithms are developed using models that are *automatically* translated into programming languages like C. These model-based engineering tools offer a higher level of abstraction closer to the problem space, e.g., the physics, therefore reducing development effort and generally increasing software quality [13]. For many years, model-based engineering has been the state of the practice in the automotive domain [14]. Unfortunately, existing metrics for manual coding are often inappropriate for auto-generated code and hence there is a need to shift the focus from code-based metrics to metrics based on models [33]. However, this topic is not fully covered in existing standards like [7].

Besides the changes on the lowest level of software development, there are also tremendous changes on the highest level of the software architecture: In the past, automotive development was ECU-centric, i.e., suppliers developed the software for an ECU either from scratch or on top of manufacturer-specific operating systems. Recent years have seen a change from these manufacturer-specific architectures towards AUTOSAR. The main goal of AUTOSAR is to enable the development of hardware-independent application functions (AUTOSAR application software) that can be executed in any AUTOSAR environment. AUTOSAR offers a basic layer that abstracts from the underlying hardware like microcontrollers and CPUs. On top of this basic layer, the AUTOSAR RTE implements a common API for the communication between application software components, regardless of them being deployed on the same ECU or on different ECUs. The next generation of AUTOSAR, which is currently under development, is called "AUTOSAR adaptive". It addresses the needs of modern cars like vehicle-to-X connections. Besides other things, AUTOSAR adaptive offers means for service-oriented architectures, which leads to a change in APIs from rather simple C-based APIs in AUTOSAR classic towards C++-based APIs describing services. Thus, while AUTOSAR fosters reuse and hence generally leads to a reduction of development effort, this comes at the price of more complex APIs and more complex toolchains, for example to synchronize the interaction between application functions and the AUTOSAR basic layer.

The 1990es and the early 2000s also saw a dramatic change in E/E architectures in the automotive domain. The 1990es were dominated by an increase of the number of ECUs from a few dozens to 100s of ECUs up to a point where managing the complexity of the overall system architecture was no longer feasible. This period was followed by a consolidation of the number of ECUs, starting with domain-based architectures [15] that organize the overall vehicle bus around more powerful (but still classical embedded) domain controllers. The last years have seen a trend towards even more consolidated ECUs. An example is Audi's Zfa (Zentrales

Fahrer-Assistenzsystem), which includes a main processor, a graphic processing unit, and a safety system – all on one integrated circuit. Those highly integrated circuits offer computational power comparable to (or exceeding) the computational power of modern business PCs. Looking at software complexity, it seems that this consolidation helps to reduce the overall complexity as more functionality is offered on the same control unit, reducing external communication. However, looking closer, these paradigm changes in E/E architectures have actually only led to a shift in software complexity: Technologies such as hypervisors (which are necessary, e.g., to tackle freedom from interference) introduce new difficulties previously not present in the automotive domain.

From an overall system perspective, we see two major trends that have a big impact on software development in the automotive domain. The first one is the trend towards connected vehicles. Today, a vehicle is no longer an isolated system. Instead, a vehicle is connected to a multitude of other systems. Modern cars offer connections to smartphones using Android Auto or Apple Carplay. Furthermore, vehicle-to-vehicle or more generically vehicle-to-X connections can be used (e.g., for intersection assistants, which enable cars to coordinate the priority at intersections among themselves). Applications for these vehicle-to-X communications are numerous and practitioners have only started to investigate the potentials of these technologies. In the past, the software of a vehicle could be seen as an isolated entity, interacting with its environment only indirectly via the laws of physics. However, with vehicle-to-X connections, the software interacts directly with the environment. This makes a vehicle *one* system in a *system of systems*, and we have to find ways of dealing with the associated complexity increase.

Another big trend is the development of highly automated and autonomous vehicles. Artificial Intelligence (AI) techniques like Machine Learning and especially Deep Neural Networks are already changing software development in the automotive domain dramatically. This change is not only about a new technology, but about how algorithms are developed and validated. Traditionally, algorithms have been completely developed and implemented by a human developer based on intermediate artifacts, such as requirements specifications[1]. In a Machine Learning setting, the major task of a human engineer is to choose the right learning engine for the task at hand and prepare enough *good examples* from which the learning algorithm can infer the desired behavior. As a consequence, the artifacts that influence the quality of a learning-based algorithm are very different from those found in conventional development. Moreover, the behavior of neural networks is still in most cases unpredictable and not understandable by humans, making uncertainty management [16] and explainable AI [17] important research topics. Ultimately, the quality of AI-based algorithms is nowadays still validated using vehicle tests, which involves a lot of manual effort. TÜV Süd experts estimate that 100 million safety scenarios have to be tested before one single fully autonomous driving function can be approved [18]. Therefore, research projects like PEGASUS [19] or VirQKo [20] investigate new engineering methodologies like virtual engineering. In virtual engineering, simulation techniques are used to test the quality of AI-based components in early stages of the classical V model. While this mitigates the validation problem of AI-based components, it adds another artifact – the simulation models – which heavily influences the overall quality of the final vehicle. Overall, new and novel approaches are required to evaluate the quality of AI-based approaches based on the newly introduced artifacts, like training data and simulation models.

Besides the aforementioned technical changes, recent years have also seen changes in development processes in the automotive domain. Agile development approaches like continuous engineering are already being implemented in industry. For example, Tesla reacts efficiently to customer requirements requested on social networks. In a particular case, a Tesla car owner sent a recommendation via Twitter directly to Tesla's CEO Elon Musk, suggesting moving the car seat back and raising the steering wheel when the car is parked. Elon Musk answered twenty-four minutes later via Twitter, saying that the feature would be included in the next software update [21]. The requested feature was released by Tesla less than two months later and deployed over-the-air. The implications of this case on engineering departments are tremendous, and the realization of such immediate responses is only possible by incorporating continuous engineering practices into development processes, from market and product monitoring to architecture analysis, redesign, verification, and deployment. The literature discusses that, in continuous software engineering, the classical V-model is no longer sequential, but is rather based on iterative executions of activities on the left side of the V (decomposition and definition) as well as on the right side (integration and validation) [22, 23]. Tesla also applies

---

[1] Note that this is also true for more traditional AI algorithms like image recognition based on edge detection.

DevOps practices: The whole fleet is running the autopilot software in the background, constantly collecting data that is shared with the company to improve the next version of the autopilot [24]. It is yet to be shown whether DevOps will have the same impact on the automotive domain as for other domains. But other agile approaches like SCRUM are already established state of the practice in the automotive domain. Without doubt, agile approaches foster faster development. However, the focus on releasing new features fast entails the risk of introducing technical debt [10], like code smells or architecture erosion. Hence managing maintainability and controlling technical debt becomes more important in an agile setting. Therefore, we need *faster feedback cycles* regarding the quality of the software in order to be able to fix problems early.

In summary, there are a lot of changes in programming paradigms, architectures, and processes going on in the automotive domain that demand novel approaches to manage the quality of software.

## 3 State of the Practice in Automotive Software Complexity Evaluation

In recent decades, the software industry has evolved a number of practices for managing software quality, including organization-specific and cross-organizational standards. As complexity has been acknowledged to be one of the main factors impacting software quality (e.g., [8]), many of the established practices aim at managing complexity in software development. These efforts include monitoring the complexity of software products, such as architecture or code, and improving software development environments, including processes, tools, and the capabilities of human resources.

Consistent findings in research and practice regarding the relationships between software process and product quality have led to efforts being initiated to establish universal guidelines for managing software quality. One of the key elements of these guidelines is managing software complexity as a critical factor influencing software quality. As a result, a number of organization-specific as well as cross-organizational guidelines and (quasi)standards have been developed. Some of them focus on software products, others on software development processes. Manufacturers, e.g., in the automotive domain, use these standards internally to manage software development and externally as a reference for agreements with their subcontractors. For example, they require their suppliers to comply with certain process maturity levels of Automotive SPICE [25] or deliver code that does not exceed the metric thresholds defined in the HIS[2] source code metrics document [7]. Common use of code metrics and coding guidelines have led to a rush of code measurement and evaluation tools, which have gradually become an inherent part of software development environments. For example, nightly build systems launch code analysis and create code quality reports automatically, or coding environments provide real-time evaluation of code compliance with coding rules. Some tools, such as Perforce's QAC, offer explicit compliance checks against coding standards such as the HIS metric set or the MISRA coding guidelines [26].

Measurement tools typically offer visualization and drill-down functionalities, which support software developers in identifying potential quality issues and their root causes quickly. Tools also typically offer different views for various quality stakeholders, for example, an aggregated view for product owners and project managers and a detailed view for developers and testers. A significant issue in practice is that not all stakeholders understand certain metrics, i.e., the concept they represent. One reason is the complexity of the metrics; another reason is the inconsistent implementation of what appear to be the same metrics by different measurement tools, yielding inconsistent measurement results. Furthermore, the impact of specific metrics on software quality is not clear and varies from context to context. In this respect, what software measurement tools are largely missing is the analysis of the actual impact that the internal characteristics of software products (e.g., complexity) have on the quality of the software in use and operation (e.g., maintainability or reliability) in a specific context. On the one hand, such capability would allow context-specific learning about quality dependencies and thus enable deriving specific improvement actions. On the other hand, modeled dependencies would allow predicting the effects of development activities on software quality.

Initial attempts have shown the great potential of quantitative data analysis – especially Machine Learning methods – in finding relevant quality patterns in development and runtime data (in the context of DevOps and DataOps). For example, in the ProDebt project [10] aimed at the quantitative analysis of software technical debt, the number of preprocessor lines of code and nesting levels were identified as critical factors

---

[2] Hersteller Initiative Software (Eng.: Manufacturer's Initiative for Software)

influencing development productivity in addition to cyclomatic complexity. The greatest challenge in applying quantitative analysis to software data is their limited quality. For example, changes to software code must be aligned with information in the project backlog and the change tracking system. However, in practice, these data sources are typically not consistent or lack important associations.

Yet, over the time, the intended strength of cross-organizational standards being universally applicable in different organizations turned out to be their greatest weakness. The "one-size-fits-all" principle proved to work only to a certain extent. For example, regarding software code, general coding guidelines proved to be useful for maintaining the quality of code and for avoiding technical debt related to maintaining and evolving software code. However, fixed thresholds defined for specific static properties of code, such as cyclomatic complexity, turned out to be challenging for the management of software quality [27–29]. Staying with the example of cyclomatic complexity, typically a high value is considered negative. However, in many cases, these values are reasonable and as such do not have any negative impact on software quality at all. One example might be model-based development, in which code is generated based on models. Because developers do not directly maintain such code, its complexity does not have an impact on the maintainability or fault-proneness of the software. Consequently, new versions of standards discard predefined threshold values for specific static properties in favor of measurable coding rules and guidelines for software processes and documents. For example, the latest MISRA compliance guidelines for C++ [26] still confirm the usefulness of code measurement in the Section "Metrics", but under the condition that it fits a specific context of a function to be implemented:

> *The use of metrics is recommended by many software process standards as a means to identify code that may require additional review […] However, the nature of the metrics being collected, and their corresponding thresholds, will be determined by the industry, organization and/or the nature of the project. This document, therefore, does not offer any guidance on software metrics.*

Hence, MISRA does not define any specific metrics or metric value thresholds. We can thus expect that the HIS consortium [7], which defined acceptance thresholds for selected code metrics, will follow the example of MISRA and refrain from predefining specific metrics and thresholds in its next version.

From the process perspective, reference models are commonly used to check the compliance of software development, thereby fostering high quality of the produced software. Examples of cross-organizational models include the Automotive SPICE process capability model [25] and the ISO 26262 [30] standard for road vehicles functional safety. Automotive SPICE includes a specific process area for measurement (MAN.6), which defines certain practices that have to be performed to set up a proper measurement program for software development projects, products, processes, and the organization. It explicitly requires that organizations derive appropriate metrics from information needs and does not propose standard measures to be collected. Regarding ISO 26262, part 6 of the standard describes recommendations for software artifacts embedded in road vehicles. One of the recommended aspects is the need to measure test coverage at both the unit and architecture level. The recommended metrics for the unit level are Statement Coverage, Branch Coverage, and Modified Condition/Decision Coverage (MC/DC). The recommended metrics for the architecture level are Function Coverage (functional completeness) and Call Coverage (functional invocation), as described in Table 15 of ISO 26262.

An example of an organization-specific standard is the Maintainability Index of Bosch [31], which is embedded in the development process and helps to define the right consequences and measures for software modules (see also chapter 4.4 Complexity of Source Code).

In general, process standards unanimously promote measurement and data analysis for managing software quality, spotting weak points, and deriving improvement actions. Similar to recent product-oriented standards, they largely advocate adapting measurement to a specific context rather than prescribing specific measurement approaches and measures (with some exceptions).

# 4 Research in Software Complexity and Quality Evaluation

In this section, we provide an overview of the current state of research in the area of software complexity and quality evaluation. We will start this overview by discussing approaches that focus on the product perspective, i.e., metrics trends in the context of different software development artifacts as well as proposed quality modeling and assessment approaches. We will close the overview with recent work motivated by the increasing relevance of AI systems. Such software systems that contain data-driven components do not only call for new complexity metrics but also for a discussion of AI-related quality characteristics.

© Fraunhofer IESE and Robert Bosch GmbH 2021        6/21        27.10.2021

## 4.1 Complexity and Quality of Requirements Specifications

Complexity measures applied to the *software requirements specification* either target or characterize qualities of the requirements itself, e.g., their understandability, or they try to provide a proxy for the complexity of the solution that is specified, e.g., its functional size. Metrics that can be used to quantify the quality-specific characteristics of requirements depend on how the requirements are specified. Metrics for natural language-based requirements comprise metrics on, e.g., vagueness and weaknesses based on indicator words [32, 33] and more advanced metrics to identify, e.g., over-specification and duplication in requirements specifications [34]. In contrast to metrics, which can be collected automatically on requirement specifications, functional size (e.g., function points) is manually derived by skilled humans following standards like IFPUG [35] or COSMIC [36].

In the context of functional sizing, Automated Function Point (AFP) was specified by the CISQ consortium, founded in 2010. AFP was developed with the objective of automating function point counting on the code level for transaction-based systems and has recently been published as the ISO/IEC 19515 standard [37]. Despite the increasing usage of transactional databases in the embedded context, the applicability of AFP may be limited to specific types of embedded software.

## 4.2 Complexity and Quality of Architectures

There exist a number of metrics with narrow measurement goals for the analysis of software architecture complexity. [38] focuses on measuring how well a software system is decoupled into small and independent modules. [39] introduces a metric that measures how tightly coupled a system is (cohesion). [40] describes industrial evidences of the strong relationship between architecture bad smells and architecture complexity.

Another common way to measure the quality of an architecture is by defining anti-patterns or so-called bad smells, which are architectural patterns that will cause some quality issues in the future. For instance, if defined APIs are not used appropriately, components have too many dependencies, or well-known architectural patterns are not followed. Most organizations use existing, well-known bad smells from the literature as a basis from which they derive specific bad smells suitable for their concrete system. Some methodologies for identifying smells are using a quantifiable pattern, e.g., thresholds of metrics that indicate a bad smell if the threshold is missed. If the architecture contains the defined pattern while satisfying its quantifiable parameters, then the system contains a user-defined bad smell.

Some approaches try to integrate individual metrics to complex indexes. For instance, the goal of the Standard Architecture Index (SAI) [40] is to calculate an Architecture Complexity Index. It is necessary to establish an explicit mapping of the smell to software quality attributes and to categorize the smells based on their scope (global, system, component, module, file, and function). Thresholds and weights need to be assigned to each smell. A threshold serves to determine whether an architecture has bad smells. The weight serves to describe the severity of the smell. The SAI provides a formula for calculating the complexity index of the architecture based on the identified smells and their weight. The higher the index, the lower the quality of the product architecture, and vice versa.

## 4.3 Complexity and Quality of Models

In the course of the further spread of *model-based development*, respective complexity metrics are also discussed in the literature, especially for widely used models like block diagrams [41, 42] or state machines. Most of the proposed metrics are either based on existing source code metrics and try to transfer their ideas to models, or they adapt ideas from graphs in other domains. The first metrics on block diagram models were proposed by Menkhaus et al. [43]. They applied McCabe's cyclomatic complexity and several OO metrics (like instability [44]) to the model graph and checked their correlation with faults. Olszewska et al. [45] defined several metrics for block diagrams, in particular size and structural complexity metrics. The latter are based on the architecture quality metrics by Card/Glass [46]. Hu et al. [47] mapped ISO 9126 maintainability quality attributes to Simulink metrics. However, a validation of this mapping is missing. Stuermer and Pohlheim [48] propose a complexity metric that is based on Halstead's software science. Dajsuren et al. [49] propose mod-ularity and complexity metrics for Simulink. They performed correlation analysis and identified several coupling metrics that are highly correlated with expert opinion on coupling. Olszewska et al. [41, 45] present a selection and (weak) evaluation of a set of previously published Simulink metrics. They found a positive correlation between structural complexity metrics and number of faults.

In summary, although a lot of studies have been performed on measuring models, there is only very limited tool support for these metrics (i.e., only MXAM / MXRAY[3] and Simulink Check[4] implement a few of them), which probably also hinders their widespread application. Furthermore, their relation to quality attributes, and in particular to maintainability, has hardly been investigated. On the other hand, complexity should be measured on the artifact that the developer works on. The code that is generated from these models might be quite complex when the models are still easy to understand, as they abstract away a lot of the details. So measuring on the model level is highly desirable for quality assessment.

### 4.4 Complexity of Source Code

Regarding *source code* metrics, in addition to the well-known metrics proposed, e.g., by Halstead [50], McCabe [51], and Chidamber and Kemerer [52], various revisions, adaptations, and alternatives are discussed in the literature [53]. In particular, recent work propagates strengthening the connection between the measured complexity and the complexity as perceived by developers when they try to understand the code. Such an approach was already proposed in the 1990s by Oman et al. [54, 55]. Using linear regression, they constructed a maintainability predictor based on expert opinions. This approach has been used successfully for maintainability assessment at Bosch [31]. Another example is Cognitive Complexity, a measure adapting McCabe's Cyclomatic Complexity metric [56]. A meta-analysis from 2020 indicates that Cognitive Complexity is well correlated with the time required to understand source code, but it does not answer the question regarding a reasonable threshold [57]. A similar idea based on cognitive weights was already published earlier [58].

### 4.5 Software Quality Standards and Quality Modeling

ISO/IEC 25000 [1] describes recommendations for Software Product Quality Requirements and Evaluation (SQuaRE) and distinguishes between three different types of interrelated qualities: Internal quality (when software is developed) has an impact on external quality (when the system is delivered to a customer), which, in turn, has an impact on software quality in use. At the core of the ISO/IEC 25010 [59] series is a definition of eight quality characteristics, further refined into 31 sub-characteristics for internal and external software quality. The standard also proposes some metrics for quantitatively evaluating quality characteristics, but the example measures provided in ISO/IEC 25023 [60] are often too abstract to be directly implemented in practice.

In order to fill the conceptual gap between abstract quality characteristics and specific metrics, quality modeling and evaluation approaches such as Quamoco [9] propose the introduction of an intermediate layer with product factors consisting of a concrete entity (artifact or part of an artifact from software development) and a relevant quality property (e.g., coupling of a software module, cognitive complexity of a source code function, or redundancy of a source code fragment). Major challenges are seen in the consistent and efficient adaptation of quality models for a specific context [29] and in providing a coherent overall picture of quality based on concrete measurement results [61]. However, even within one application domain, expert opinion differs regarding what quality aspects are important and what evaluation rules to use. For this reason, the practical adaptation of these approaches is currently limited.

Vogel et al. [28] present a systematic literature review on metrics in automotive software development and their mapping to ISO/IEC 25010. One of their findings is that out of a total of 112 metrics, only very few come with boundary values, because boundary values are seen as specific to particular contexts and project setups. Furthermore, the majority of metrics focus on maintainability as a quality characteristic and code as an analysis object. Broad tool support for a lot of the metrics listed is limited.

---

[3] https://model-engineers.com
[4] https://www.mathworks.com

### 4.6 Quality and Technical Debt

Technical debt is a concept closely related to quality and refers to the cost implied by additional work caused by choosing an easy, yet low-quality, solution now instead of using a proper approach that would take more effort now but allow avoiding additional costs later. Technical debt is an inherent phenomenon of evolving software; as software is continuously changed, its structure deteriorates, which is manifested in growing complexity. Measuring and managing technical debt were important research topics in the last decade, since technical debt allows representing the implications of insufficiently maintainable software in monetary units. An important challenge in calculating technical debt is that highly simplified cost models lead to unrealistic cost estimates, while in many cases the costs associated with individual deficits cannot be calculated since there is a lake of data that would provide traceability between a specific deficit and related maintenance efforts [10, 62]. [63] discusses the usage of software quality evaluation in projects with short development cycles, and [64] propose adaptations for the context of DevOps.

Regarding architectural debt, [65] presents an industrial case study and a cost-benefit model for estimating possible savings. [66] presents a use case in which an economic model was used to assess the effort that could be saved after refactoring the architectural debts.

### 4.7 Quality and Artificial Intelligence

Systems are increasingly highly automated and intelligent. Artificial Intelligence and Machine Learning, in particular, are increasingly proving to be important enablers for automation and autonomy. However, guaranteeing and verifying essential properties such as trustworthiness and dependability of ML components is still a largely unsolved challenge. Therefore, Machine Learning outcomes suffer from various kinds of uncertainties [67]. Due to the different nature of ML, we have to re-interpret existing qualities (e.g., from ISO/IEC 25010) for ML systems or add new ones (such as trustworthiness). For instance, [12] provides an overview of the quality properties that are typically relevant for an entity of an ML component of interest (such as completeness of training data or correctness of the trained model) and discuss how to objectively evaluate adherence to quality requirements.

## 5 Needs for Action

Whereas in research, many new approaches have evolved for coping with the trends in automotive software development, in practice, we find a quite traditional, largely static view of software quality in general and its complexity in particular. For instance, it is quite common for manufacturers to demand that their suppliers deliver a list of complexity measures (based on the HIS) with static thresholds to fulfill. This approach neither considers the fact that software quality is dependent on the specific context of a function to be implemented, nor does it address the trends sketched earlier. We therefore propose changing the paradigm of measurement-based software quality evaluation to a more modern and holistic approach, including conclusions from recent research and addressing novel technologies and development processes.

We see the following needs and requirements for future measurement-based software quality evaluation:

**Consider the perspectives of different stakeholders:** In modern software development, there are different stakeholders with more long-term or more short-term interest in software quality and different values related to software quality [68]. While the core functionality of vehicles, namely getting securely and comfortably from point A to point B, becomes a commodity, differentiation among manufacturers is more about software and its usability and reliability. The end user in a connected car with assistance and multimedia functions is confronted much more directly with software and its quality. From a development perspective, we have the strong interest to update software functionality more frequently. This means that the term software quality must be sketched and analyzed from the perspectives of all relevant stakeholders (such as end users, product managers, project managers, developers, etc.).

**Evaluate relevant development artifacts:** Traditionally, the focus of measurement-based quality evaluation has mostly been on the code level. Considering modern development paradigms, this evaluation scope must be broadened. Nowadays, it is much more about model-based development, where code is partially generated automatically and the overall architecture of the software system becomes more important. The inclusion of new types of components built with new technologies, e.g., AI-based components, also requires the inclusion of related development artifacts other than source code, e.g., Machine Learning models and the data on which the models were trained.

**Go beyond the software component level:** Traditionally, the focus of measurement-based quality evaluation has mostly been on the level of single software components (functions/methods and their modules/classes). Nowadays, higher levels of the system architecture are much more relevant from a quality perspective. The overall software architecture goes beyond the level of a single control unit. For instance, we find whole software architectures for different types of controllers, for the vehicle as a whole, for a network of connected vehicles up to the level of digital platforms and mobility services. Therefore, we must also broaden the scope and go beyond the component level in order to better understand the effect of software quality for different stakeholders.

**Evaluate relevant quality properties:** Traditionally, the focus of measurement-based quality evaluation has mostly been on quite technical quality properties, especially on complexity, because these are easy to measure automatically. Nowadays, software quality is about much more than size and cyclomatic complexity. In fact, the field of potentially relevant quality properties to consider is so large that we must have a clear rationale for deciding what qualities are relevant for the specific goals of a particular software in a dedicated context. For instance, for a connected vehicle, data security and privacy also become very important issues. For automated driving or assistant functions based on AI, trustworthiness (or sub-aspects such as explainability) of the AI model used become important. The focus should be on qualities and corresponding metrics that are really useful for stakeholders in the project, and especially for developers. So, the aim should not be to cover all potential qualities and related metrics, but to concentrate on the most important ones (less is more). Those relevant qualities and metrics should have high usability and allow addressing the real problems / issues regarding the software.

**Be more flexible and less static:** Traditionally, the metrics collected have mostly been static with fixed thresholds. This assumes a universal generic understanding of which metrics to collect for all kinds of systems and universal thresholds defining what high-quality software systems must look like. Maybe decades ago, ECU software was not as diverse as today and it was a valid attempt to come up with a fixed set of metrics and thresholds for determining its quality. Nowadays, software quality could mean quite different things for different kinds of systems/functions and stakeholders in the context of automotive software development. Furthermore, the notion of software quality may change over time. Measurement-based quality evaluation should not impede development flexibility and speed. This especially means that the specification of software quality is not something that is only defined once, but must also be allowed to change and be adapted over time.

## 6   Quality Evaluation Framework for Automotive Software Development

As we have seen in the previous section, there is a need to change our views on the term software quality and its evaluation. We have to move from a mostly static set of complexity metrics with fixed thresholds towards a more flexible approach that considers the context as well as goal-specific requirements for managing software quality.

> The objective of the framework is to provide a systematic approach for defining what the term software quality means in a goal-oriented way for a specific context of a function to be implemented, and to measure and evaluate relevant qualities systematically in order to enable early identification of potential quality risks and issues in the course of a software development and maintenance project.

The framework consists of four areas:

1. Understand the context and the stakeholder goals/needs
2. Derive technical quality factors for relevant development artifacts related to stakeholder goals
3. Determine quality measures and evaluation rules for relevant quality factors
4. Define a process for managing quality during the entire lifecycle

The different activities and outputs of the framework operationalize a goal-oriented measurement process (such as GQM [69]), which has been a quasi-standard in measurement research for decades and is also demanded by different process maturity models (such as in the "measurement process" as defined by SPICE [5, 6]). Figure 1 gives an overview of the proposed framework.

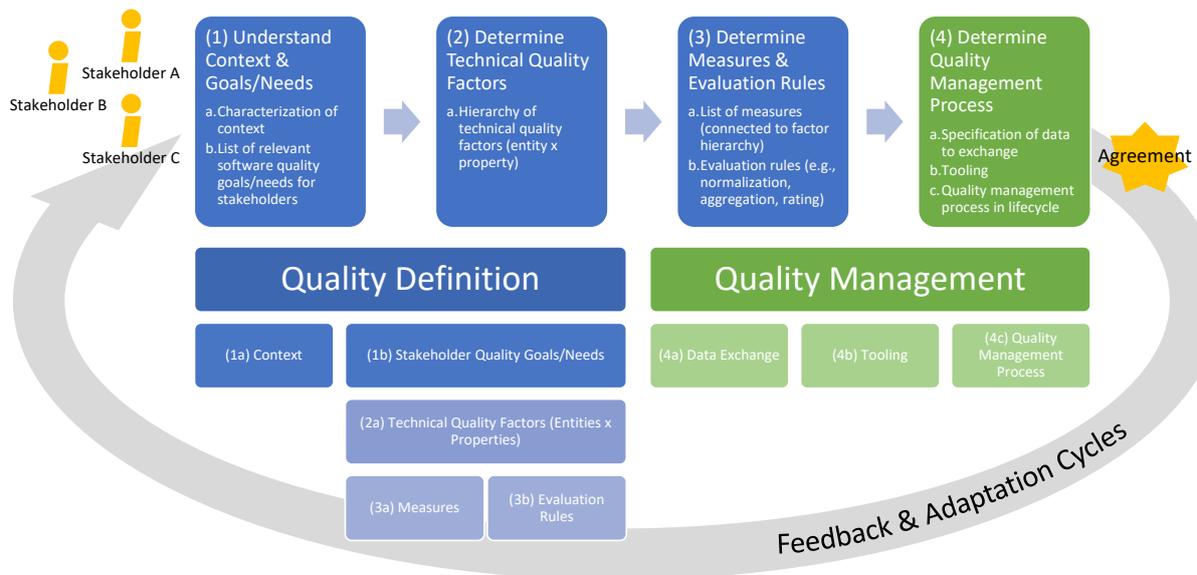

*Figure 1: Framework for Measurement-based Quality Evaluation for Automotive Software Development*

We propose that all involved stakeholders and parties should agree on goals, factors, measures, evaluation rules, as well as quality management processes at the beginning of a software development project, and that should revise that agreement in predefined cycles based on the course of the project. The agreement may also contain exceptions or special evaluation rules for dedicated parts of the software (such as off-the-shelf platforms or open source software used).

We see two options for using the framework in the context of an OEM / supplier relationship:

(a) In the first option, the agreement about goals, factors, measures, evaluation rules, as well as quality management processes would directly become part of a contract between the OEM and the supplier. The advantage of this option is that the OEM and the supplier have a highly objective evaluation basis, which is systematically derived from the goals and needs of the software product to be used in a specific context. The disadvantage is that this option is less flexible because updating and adapting the agreement in the course of the project would imply changing the contract. Furthermore, the overhead on both sides is higher because the fulfillment of the agreement must be formally proven.

(b) In the second option, only the framework for measurement-based quality evaluation would become part of a contract between the OEM and the supplier. Similar as currently done for requesting and checking a certain capability level of processes according to the Automotive SPICE model, the OEM could just request and check that the activities of the framework are performed properly and that the corresponding outputs are produced. The advantage of this option is that the OEM and the supplier can define the agreement (about goals, factors, measures, evaluation rules, as well as quality management processes) based on quality needs instead of contractual constraints. Furthermore, this allows for high flexibility in updating and adapting the agreement in the course of the project. The disadvantage lies in less formal control over the contents of the agreement, as it would not be part of the contract.

In the following sections, we will describe different aspects of the framework and give advice on suitable processes, methods, and tools.

### 6.1 Understand Context and Stakeholder Goals/Needs

This area of the framework is about characterizing the context of a function to be implemented and to discuss relevant software quality goals/needs together with all relevant stakeholders of the customer and their software supplier.

To identify relevant measures for quality evaluation, understanding the context and how the software will be used in this context is essential. For instance, for an infotainment system, the user interface and its usability might be important, whereas this would be out-of-scope for an embedded engine control unit. Furthermore, different stakeholders from different parties are typically involved in software development with different interests in software product qualities. For instance, a product manager on the customer side may be interested in the usability of a user interface, whereas a product manager of the development organization may want to be able to quickly change and adapt the software to user/customer needs (i.e., they are interested in the maintainability of the software). As an alternative to agreeing on stakeholder goals, a risk-driven approach can be used. Finally, it is important to analyze the real value that software quality provides to the different stakeholders and parties involved. Table 1 lists some examples of typical context characteristics, while Table 2 lists some examples of typical stakeholder goals related to software quality.

Note that we would recommend focusing and agreeing on the most important characteristics and goals for software quality instead of on all potential ones. Furthermore, we would recommend writing down a rationale explaining why a certain stakeholder goal was agreed upon. For instance, if frequent feature updates are planned, software maintainability is of high importance.

*Table 1: Examples of Context Characteristics*

| Product | Development Approach |
|---|---|
| • Type = Infotainment software<br>• User interface = Touchscreen<br>• Contains AI/ML components = No<br>• Architecture = AUTOSAR | • Process = Scrum<br>• Release cycles = Twice a year<br>• OTA updates = Yes |

*Table 2: Examples of Quality Goals for Different Stakeholders (based on ISO/IEC 25010)*

| Software Supplier / Customer | Customer / User |
|---|---|
| • Functional suitability<br>• Performance efficiency<br>• Compatibility<br>• Usability<br>• Reliability<br>• Security<br>• Maintainability<br>• Portability | • Effectiveness<br>• Efficiency<br>• Satisfaction<br>• Freedom from risk<br>• Flexibility |

### 6.2   Determine Technical Quality Factors

This area of the framework deals with determining technical quality factors for all relevant development artifacts related to stakeholder goals.

In order to be able to manage software quality on a technical level, abstract qualities (such as those defined by ISO/IEC 25000) must be decomposed into concrete technical quality factors that are specific for concrete usage scenarios. For instance, what does the usability of a user interface or the maintainability of software mean? To precisely specify quality-centered usage scenarios, we recommend using the approach proposed by [11], who claim that such specifications should be done in terms of: (i) the environment or condition in which this driver occurs; (ii) the event that stimulates the occurrence of the driver; (iii) the expected response of the system to the driver event; and (iv) the quantifications associated with the three previous aspects. The usage scenario specifications should be precise enough to enable the architecture to properly reason about adequate architecture solutions for addressing them.

A technical factor is specified by a quality property of a development artifact or a part of a development artifact (such as the understandability of the source code). A technical factor refines stakeholder quality goals on a technical level (e.g., the quality goal "maintainability of software" can be refined by "understandability of source code"). While the quality goals give a clear motivation of why it is important to deal with different aspects of software quality, the technical factors break that down into what entities (development artifacts or parts thereof) are of interest and what properties have to be evaluated. We would recommend starting by thinking about relevant entities first, and then identifying relevant quality properties per entity. The tricky part lies in focusing on the most important entities and their properties for which quantitative quality management must be established, instead of looking at all possible combinations. For this, it might be helpful to focus on entities that are exchanged between the parties / stakeholders and which have a high risk of leading to serious issues during development or operation time of the software to be developed.

Table 3 presents some examples of technical factors.

We would recommend describing the rationales for why a certain technical factor is related to a specific stakeholder goal. More specifically, we recommend specifying this rationale by means of pros, cons, and risks (i.e., the points in favor and against), assumptions and quantifications (i.e., the assumptions made about the decision and their related measurable effects), and trade-offs (i.e., aspects negatively impacted by this decision).

For practical reasons, it might be necessary to create a hierarchy of technical factors if a property or entity is still too abstract to be understood in the same way by all stakeholders. For instance, "understandability of source code" could be broken down to "documentation degree of public interfaces in source code" or "complexity of APIs in source code", etc. At the end, technical factors should be as concrete as possible.

*Table 3: Examples of Technical Quality Factors related to Different Development Artifacts*

| Requirements | Architecture | Source Code | Test |
|---|---|---|---|
| • Volatility of features<br>• Appropriateness of requirements breakdown<br>• Completeness and consistency of requirements acceptance criteria<br>• Traceability between requirements and architecture | • Solution adequacy of architecture to the associated requirements (ATAM)<br>• Complexity of architecture<br>• Documentation adequacy of architecture<br>• Compliance between architecture and code<br>• Traceability between architecture and source code | • Complexity of APIs in source code<br>• Algorithmic complexity of source code<br>• Coupling / cohesion between source code components<br>• Understandability of source code<br>• Documentation degree of public interfaces in source code<br>• Compliance with MISRA C/C++ coding guidelines | • Coverage of code with test cases<br>• Coverage of requirements with test cases |

### 6.2.1 Identification of Relevant Entities

Relevant entities (development artifacts or parts thereof) depend on the context and on the stakeholders' quality goals. For instance, for a domain controller, a hardware layer and its qualities are more important than for an infotainment system. For a Machine Learning (ML) component of a software, the quality of the data used for training and testing the model might by essential, while it might not be important at all for a classically programmed piece of software.

Typical entities to consider at this high level include, but are not limited to, requirements documentation, system and software architecture, and design, source code, or V&V-related artifacts. However, entities may also be specific to certain development techniques used (such as model-based development, virtual engineering, simulation) or types of systems under development (such as a system including AI components).

For instance, in the case of model-based development, SysML or MATAB/Simulink models may be of interest, and source code generated automatically from the models should be excluded. For ML components, one might include qualities related to the data, the learned model, the infrastructure in which the model is learned, the system into which the ML component is placed, or the environment in which the model is used later on [12].

### 6.2.2 Identification of Relevant Quality Properties

For each relevant entity, relevant quality properties have to be identified. This could include quality properties relevant for development time (such as reliability) or during the use of the software (such as efficiency). ISO/IEC 25010 gives some starting points for thinking about generic qualities, which could be relevant on the level of the software system. However, in order to be practically applicable, they have to be broken down into more technical properties by using approaches such as GQM [69], quality modeling approaches [9] from the field of software measurement, or approaches generically applicable for breaking down abstract concepts for decision-making, such as Multiple Criteria Decision Making [70]. One may also follow predefined methods for certain types of entities to define relevant quality properties, such as the RATE method for architecture assessment, or derive relevant qualities from architecture scenarios or drivers [11].

After relevant quality properties have been identified for all relevant entities, an analysis should be performed of potential cross-links between the created technical quality factors. This is useful for identifying potential conflicts between factors or factors that support each other. It might also be helpful to note any known relationship to relevant qualities not related to the software.

### 6.3 Determine Measures and Evaluation Rules

This area of the framework determines quality measures and evaluation rules for all relevant technical quality factors.

In order to quantitatively evaluate software quality, the technical factors must be made measurable by defining concrete metrics and clear rules on how to evaluate the measurement data and distinguish between acceptable and unacceptable levels of quality. For instance, you may define a measure for the technical factor "Documentation degree of public interfaces in source code" by "percentage of public interfaces containing a description of functionality, input parameters, and return values". A simple evaluation rule could be that "100% of public interfaces in source code must be documented".

We would recommend starting by thinking about relevant measures for all identified technical factors first, and then defining evaluation rules to identify relevant quality properties per entity.

#### 6.3.1 Identification of Relevant Measures

To identify relevant measures, approaches such as GQM may be helpful, as they support the derivation of metrics via measurement goals and questions. However, it is important not to reinvent the wheel, i.e., not to start on the green field when it comes to defining measures, but to refer to well-known measures from the literature and to use standardized measures whenever possible. On the one hand, one should avoid determining measures, which are hard or nearly impossible to collect in practice. On the other hand, one should also not only focus on those measures that are the easiest to collect, but rather on those that provide a good measure for the relevant technical quality factors.

For most standard technical quality factors, it should be possible to find recommended measures or whole evaluation models in the literature. For instance, different qualities of a software architecture could be measured using the Software Architecture Complexity Model (SACM) [71]. For different source code qualities, one also finds metric collections in the literature or collected by tool vendors (e.g., [72]), or models for addressing more complex concepts, such as Technical Debt [10]. For certain types of systems, such as AI systems, different proposals for measures to be collected for dedicated quality properties can also be found [12].

Note that there is unfortunately no one-size-fits-all solution regarding measures that should be collected. For certain technical quality factors, there exist some typical suspects. However, the number of quality factors and related measures is far too high to simply try to collect all of them. Therefore, it is crucial to know what technical quality factors are relevant to evaluate in the course of a software development project.

Note also that the main goal of the proposed framework is to describe an approach for identifying relevant metrics and not to define a standard set of metrics. In practice, external benchmarking databases, such as the TIOBE Index [73], can also be found. These benchmarking databases use a common set of measures to evaluate software systems (grouped by programming language), but do so independent of context and quality goals. This approach fits the purpose of external benchmarking, but not the purpose of focusing on all quality risks and issues that are relevant for the functionality of a dedicated software system.

#### 6.3.2 Definition of Evaluation Rules

Defining clear and meaningful evaluation rules is probably the most important task, but also challenging in the context of evaluating software quality.

Before starting to define evaluation rules, it should be clarified on what abstraction level measurement-based quality evaluation is required. So far, the described framework has specified, three different levels: (1) Stakeholder quality goals refer to a set of (2) technical quality factors, which in turn refer to (3) concrete measures. In the simplest case, a quality evaluation on the level of single measures is sufficient. However, depending on the stakeholders and their interest, it could be required that the evaluation on the level of single metrics is aggregated across the tree levels based on the defined tree/graph structure for calculating quality indexes for quality factors, quality goals, or the software as a whole. If this kind of aggregation is required, measurement values typically have to be normalized and mapped to a unique evaluation scale using, e.g., utility functions, before they are aggregated. Furthermore, the type of aggregation has to be defined, e.g., whether a compensatory aggregation model (like weighted sum) or a not-compensatory model (like worst-case aggregation) is applied. [9, 70] describe this procedure.

A critical question is how to define evaluation rules for measurement values; specifically, how to come up with meaningful thresholds for distinguishing between acceptable and unacceptable measurement values. Older approaches assume that one can define universal thresholds that are valid for a certain measure, independent of the nature and functionality of the system; for instance, assuming that McCabe's Cyclomatic Complexity (MCC) must always be lower than 10. More recent approaches assume that acceptable thresholds should be defined for a certain type of system using a more baseline-driven approach. That is, the actual distributions of measurement values are analyzed for an already existing system, or for a similar system if the system should be newly developed. Based on this baseline, realistic thresholds are defined relative to this baseline. For instance, MCC for the whole system or certain critical parts should not become higher or should be reduced by 10%.

When defining the evaluation rules, it is also important to consider which level of granularity according to the structure of the software system one is referring to. For instance, are we referring to the average or the maximum of MCC across all methods/functions of a system or to the absolute sum of MCC across the whole system, or do we use another measure for normalization, such as the size of a function measured by LOC or number of statements?

### 6.4    Determine Quality Management Process

This area of the framework defines the agreed process for managing quality during the entire lifecycle of the software product. Nowadays, a more continuous and flexible approach is required for managing quality during the entire lifecycle, which allows for a more incremental analysis of quality based on Agile principles.

First, based on the specification of software quality from the previous framework areas, all parties have to agree on the following aspects:

1. What measurement data is exchanged on which level of granularity? For instance, is raw data regarding all defined measures exchanged or just the evaluation results of all defined rules?
2. What tools are used for collecting, storing, and analyzing the measurement data, and where is the analysis performed? For instance, is the analysis performed on the supplier side, the customer side, or by a neutral party?
3. When to analyze the data, what to do if the evaluation reveals quality issues, and how to integrate measurement-based quality evaluation into the development process? For instance, are software deliveries with quality issues not allowed at all or have all issues been documented as tasks in an Agile product backlog to be addressed in future software releases?
4. When to revise the quality specification? For instance, if new quality issues are raised or critical functionality of the software system is changed, all parties should agree on adapting the quality specification.

Regarding the point in time when certain measurement data becomes available, the use of virtual engineering and simulation makes sense for early quality evaluation (e.g., virtual hardware prototypes).

Note that for certain types of systems, such as AI systems, feedback loops from operation to development also have to be considered in order to detect quality issues (e.g., by monitoring components of AI decisions).

It is also important to note that the defined process for measurement-based quality evaluation should not impede development flexibility and speed, but should become part of the process (e.g., certain Scrum reviews or retrospective meetings). This specifically includes that the specification of software quality is not something that is only defined once in the contract at the beginning of the project, but is also allowed to change in the course of the project.

## 7   Conclusions

In this paper, we discussed relevant trends in automotive software development that have an impact on software quality and how to evaluate and measure it. This included especially the increased importance of software in the automotive domain, the heterogeneity of the application fields, novel software architectures and model-based development, and changes in the development process towards more agile processes and BizDevOps. We also discussed the current state of the practice of measurement-based complexity evaluation in automotive software development. We concluded that we still find largely static approaches for measuring software quality in general and complexity in particular, which do not address the trends that we have observed in recent years. Furthermore, we highlighted some relevant approaches from research in software complexity and quality evaluation. We concluded that several new approaches have emerged in recent years for coping with the identified trends. These especially include what development artifacts to evaluate, what metrics to collect, and how to analyze, evaluate, and manage software quality.

Based on that, we summarized the main needs for action for measurement-based software quality evaluation and management from our point of view and proposed a quality evaluation framework for automotive software development that addresses the trends and progression in research. The framework aims at defining the term software quality in a goal-oriented way for a specific context and at describing how to measure and evaluate relevant qualities. Following this approach, we proclaim the following advantages when fully implementing the framework:

- Software quality has a clear link to the value it provides for all relevant stakeholders in specific application contexts.
- Only relevant technical quality factors and measures are included, which reduces the overhead for collecting measurement data that is not actually of use.
- Application- and goal-oriented evaluation rules avoid comparing apples and oranges and focus on meaningful quality risks and issues discovered.
- All relevant development artifacts and parts thereof are included for which evaluating the quality during the course of the project is important, which better reflects modern development paradigms.
- A real added value for collecting measurement data during the product lifecycle can be provided to actually address critical quality risks and issues.
- The framework allows for keeping up with increasing development speed and more flexible development approaches (such as Agile development or DevOps).

Currently, the framework is only defined on a conceptual level based on our knowledge and experience from research and practice. For detailing, discussing, and deploying the framework, we suggest the following steps as future work:

- More detailed descriptions of the sketched framework (activities, work products, roles, tools, comprehensive example, templates, etc.) have to be elaborated and provided.
- Pilot case studies where the framework is tried out on practical examples have to be performed in the context of real development projects (including getting feedback from practitioners).
- A discussion has to be initiated among relevant partners and an agreement has to be reached for using the framework in the future (e.g., between key OEMs, suppliers, and associations).
- A reference standards document has to be created based on the description of the framework.
- The standard has to be published and support for deployment and rollout has to be provided.

Based on the sketched trends regarding the importance of software in the automotive domain, we see a clear need to change the status quo regarding the evaluation and management of software quality. We see these steps as a practical way to start discussions about what a novel approach addressing the trends and progression in research could look like.

## Acknowledgments

The authors would like to thank Robert Bosch GmbH for supporting the creation of this position paper. Furthermore, we would like to thank Andreas Jedlitschka, Ralf Kalmar, Sonnhild Namingha, and Hendrik Post for their valuable comments and for reviewing initial versions of the paper. We also want to emphasize that this article represents our own independent position on the topic based on our personal experience and knowledge.